\documentclass[aip,cha,reprint,onecolumn]{revtex4-1}

\usepackage{graphicx}
\usepackage{epsfig}
\usepackage{subfigure}

\usepackage{amsmath,amsthm,amsfonts,latexsym}

\begin{document}

% Use the \preprint command to place your local institutional report
% number in the upper righthand corner of the title page in preprint mode.
% Multiple \preprint commands are allowed.
% Use the 'preprintnumbers' class option to override journal defaults
% to display numbers if necessary
%\preprint{}

%Title of paper
\title{Theoretical foundations of emergent constraints: relationships between climate sensitivity and global temperature variability in conceptual models}
% repeat the \author .. \affiliation  etc. as needed
% \email, \thanks, \homepage, \altaffiliation all apply to the current
% author. Explanatory text should go in the []'s, actual e-mail
% address or url should go in the {}'s for \email and \homepage.
% Please use the appropriate macro foreach each type of information

% \affiliation command applies to all authors since the last
% \affiliation command. The \affiliation command should follow the
% other information
% \affiliation can be followed by \email, \homepage, \thanks as well.
\author{Mark S. Williamson}\email{m.s.williamson@exeter.ac.uk}
\affiliation{Exeter Climate Systems, College of Engineering, Mathematics and Physical Sciences, University of Exeter, Laver Building, North Park Road, Exeter EX4 4QE, UK}
\author{Peter M. Cox}
\affiliation{Exeter Climate Systems, College of Engineering, Mathematics and Physical Sciences, University of Exeter, Laver Building, North Park Road, Exeter EX4 4QE, UK}
\author{Femke J. M. M. Nijsse}
\affiliation{Exeter Climate Systems, College of Engineering, Mathematics and Physical Sciences, University of Exeter, Laver Building, North Park Road, Exeter EX4 4QE, UK}
%Collaboration name if desired (requires use of superscriptaddress
%option in \documentclass). \noaffiliation is required (may also be
%used with the \author command).
%\collaboration can be followed by \email, \homepage, \thanks as well.
%\collaboration{}
%\noaffiliation

\date{\today}

\begin{abstract}
\begin{description}
\item[Background] The emergent constraint approach has received interest recently as a way of utilizing multi-model General Circulation Model (GCM) ensembles to identify relationships between observable variations of climate and future projections of climate change. These relationships, in combination with observations of the real climate system, can be used to infer an \emph{emergent constraint} on the strength of that future projection in the real system. However, there is as yet no theoretical framework to guide the search for emergent constraints. As a result, there are significant risks that indiscriminate data-mining of the multidimensional outputs from GCMs could lead to spurious correlations and less than robust constraints on future changes. To mitigate against this risk, Cox et al \cite{ref:Cox18} (hereafter CHW18) proposed a theory-motivated emergent constraint, using the one-box Hasselmann model to identify a linear relationship between equilibrium climate sensitivity (ECS) and a metric of global temperature variability involving both temperature standard deviation and autocorrelation ($\Psi$). A number of doubts have been raised about this approach, some concerning the application of the one-box model to understand relationships in complex GCMs which are known to have more than the single characteristic timescale.

\item[Objectives] To study whether the linear $\Psi$-ECS proportionality in CHW18 is an artefact of the one-box model. More precisely we ask `Does the linear $\Psi$-ECS relationship feature in the more complex and realistic two-box and diffusion models?'.

\item[Methods] We solve the two-box and diffusion models to find relationships between ECS and $\Psi$. These models are forced continually with white noise parameterizing internal variability. The resulting analytical relations are essentially fluctuation-dissipation theorems.

\item[Results] We show that the linear $\Psi$-ECS proportionality in the one-box model is \emph{not} generally true in the two-box and diffusion models. However, the linear proportionality \emph{is} a very good approximation for parameter ranges applicable to the current state-of-the-art CMIP5 climate models. This is not obvious - due to structural differences between the conceptual models, their predictions also differ. For example, the two-box and diffusion, unlike the one-box model, can reproduce the long term transient behaviour of the CMIP5 \texttt{abrupt4xCO2} and \texttt{1pcCO2} simulations. Each of the conceptual models also predict different power spectra with only the diffusion model's pink $1/f$ spectrum being compatible with observations and GCMs. We also show that the theoretically predicted $\Psi$-ECS relationship exists in the \texttt{piControl} as well as \texttt{historical} CMIP5 experiments and that the differing gradients of the proportionality are inversely related to the effective forcing in that experiment.

\item[Conclusions] We argue that emergent constraints should ideally be derived by such theory-driven hypothesis testing, in part to protect against spurious correlations from blind data-mining but mainly to aid understanding. In this approach, an underlying model is proposed, the model is used to predict a potential emergent relationship between an observable and an unknown future projection, and the hypothesised emergent relationship is tested against an ensemble of GCMs.

\end{description}
\end{abstract}

%\pacs{03.67.Mn, 03.65.Vf, 11.15.Ha, 11.30.Cp}
\maketitle

\section{Introduction}

%Explain approach:
Emergent constraints \cite{ref:Allen&Ingram02,ref:Hall&Qu06} provide a promising way to relate observations of the present day to future projections of the climate. The usual approach is to take a model ensemble (such as the multi-model CMIP5 ensemble \cite{ref:CMIP5}) and use it to find a relationship via a scatter plot between an observable plotted on the $x$ axis and the future projection plotted on the $y$ axis, each point on the plot being one member of the model ensemble. The model ensemble derived relationship or \emph{emergent relationship}, and the uncertainty in it, can then be determined from regression on the scatter plot. A measurement of the observable in the real world can be combined with the model-derived emergent relationship to produce an \emph{emergent constraint} on the climate projection.

%Previous literature:
Hall and Qu \cite{ref:Hall&Qu06} published one of the first emergent constraints relating the strength of the snow albedo feedback in the seasonal cycle (the observable) to the strength of the snow albedo feedback in climate projections within the multi-model ensemble used in IPCC AR4 \cite{ref:IPCC_AR4}. Since then many others have been published including studies on sea-ice\cite{ref:Boe09,ref:Massonnet12}, tropical precipitation extremes\cite{ref:OGorman12}, equilibrium climate sensitivity (ECS)\cite{ref:Annan&Hargreaves06,ref:Huber10,ref:Fasullo&Trenberth12,ref:Brown&Caldeira17,ref:Cox18}, carbon loss from tropical land under warming\cite{ref:Cox13}, zonal shift of Southern Hemisphere westerlies\cite{ref:Kidston&Gerber10}, cloud feedbacks \cite{ref:Brient&Bony13,ref:Sherwood14,ref:Klein&Hall15, ref:Tian15,ref:Tsushima16,ref:Lipat17}, strengthening of the hydrological cycle\cite{ref:DeAngelis15}, the climate-carbon cycle feedback\cite{ref:Wenzel14} and CO$_2$ fertilization effect\cite{ref:Wenzel16}, future changes in ocean net primary production\cite{ref:Kwiatkowski17}, permafrost melt\cite{ref:Chadburn17}, and changes in natural sources and sinks of CO$_2$\cite{ref:Hoffman14}.

%Limitations of the approach:
Some scepticism about emergent constraints is healthy, particularly when they are not founded on well understood physical processes. There are significant risks that indiscriminate data-mining of the multidimensional outputs from models could lead to spurious correlations \cite{ref:Caldwell14} and less than robust constraints on future changes \cite{ref:Bracegirdle13}. Care is also needed drawing statistical inferences from ensembles of small numbers of models. The problem is compounded if models within the ensemble share common components giving a smaller effective ensemble size \cite{ref:Pennell11,ref:Masson&Knutti11,ref:Herger17}. Observations used to guide model development also may lead to dependencies \cite{ref:Masson&Knutti12}.

% What we do here, re-write these when I know what I'm talking about
To minimise these risks, a theoretical framework for finding and evaluating emergent constraints is needed. The approach described here involves a form of hypothesis testing, in which an underlying simple, conceptual model is proposed, the model is used to predict an emergent relationship between an observable and an unknown future projection, and the predicted emergent relationship is tested against results from an ensemble of more complex models. Emergent relationships are usually assumed to be univariate and linear, but these are not necessary simplifications. As an example, we illustrate this theory led approach using simple conceptual models of the global mean temperature as emergent constraints on ECS and test the theoretically predicted relations against observations and the CMIP5 models.

In CHW18, the theoretical linear relationship between a measure of the variability of annual mean global surface air temperature, the observable $\Psi$, and the equilibrium climate sensitivity (the future projection) was used to derive an emergent constraint on ECS. Colman and Power\cite{ref:Colman&Power18} also found a correlation between the tropical decadal temperature standard deviation and ECS in the CMIP5 models. A number of doubts have been raised about CHW18\cite{ref:Brown18,ref:Po-Chedley18,ref:Rypdal18,ref:Cox18a}, some concerning the theory and the application of the one-box model to understand relationships in complex GCMs which are known to have more than the single characteristic timescale\cite{ref:MacMynowski11,ref:Geoffroy13}. In section \ref{sec:theory} we investigate whether the relation in CHW18 derived for the one-box model still holds in more realistic yet still analytically soluble conceptual models, namely the often used two-box and diffusion models. It is known the two-box and diffusion models unlike the one-box model are able to reproduce the long term transient behaviour of the CMIP5 GCM \texttt{abrupt4xCO2} and \texttt{1pcCO2} simulations\cite{ref:Geoffroy13,ref:Caldeira&Myhrvold13}. Although we find the one-box linear proportionality between $\Psi$ and ECS is generally no longer true in the two-box and diffusion models, we show the linear proportionality holds to a good approximation for both when the range of their parameters are applicable to the complex CMIP5 GCMs. This gives us increased confidence in the theoretical foundation of CHW18.

It is important to note that each of these conceptual models differ structurally, predict different temperature responses and will not be able to reproduce all of the features of the global mean temperature response.  One could loosely view these conceptual models as zeroth order approximations and GCMs as higher order approximations of the real world. The often used quote `all models are wrong but some are useful' is quite apt as a guiding principle for this manuscript. The usefulness of the model will depend on the question asked of it.

In section \ref{sec:CMIP5comp} conceptual model predictions are compared with the CMIP5 ensemble and observations, particularly the power spectra and autocorrelation functions. The pink power spectrum of global mean temperature in observations and CMIP5 models can only be reproduced by the diffusion model. However, if one is interested in the shorter time scale behaviour for use as an emergent constraint on ECS, we find the simplest conceptual one-box model will serve as a good approximation.

Also in section \ref{sec:CMIP5comp} $\Psi$ vs ECS emergent relationships for both \texttt{piControl} and \texttt{historical} CMIP5 experiments are shown. Both have the theoretically predicted linear proportionality although they have differing gradients. This difference in gradient is theoretically expected to scale inversely with the effective forcing and this is also observed in the CMIP5 models.

The current paper can be seen as a companion paper to CHW18, as it examines and tests the appropriateness of the theory used to inform that study.
%
%Other references: Radiation budget constraint on ECS (CMIP3) \cite{ref:Huber10}, review of emergent constraints for cloud feedbacks \cite{ref:Klein&Hall15} (they also discuss three classes of emergent constraint and how they should be judged). \cite{ref:Caldwell15} not emergent constraints but on the sources of intermodel spread in ECS (reason clouds). Changing climate sensitivity with time from regional warming patterns \cite{ref:Armour12}. .

%\section{Hypothesis}

%The approach we advocate here is a form of hypothesis testing in which a simple, physically justified model is proposed that relates the observable to the future projection, preferably quantitative but at least giving the functional form. For example, is the observable a linear, quadratic or more exotic function of the future projection? This simple model will of course not be perfect but should should give a useful prediction of this functional form that can be tested against members of the model ensemble.

%In the next section we use three simple models to relate statistics of the annual global mean surface air temperature anomaly ($\Delta T$) to the equilibrium climate sensitivity (ECS).

\section{Conceptual models relating global temperature variability to equilibrium climate sensitivity}\label{sec:theory}

Caldeira and Myhrvold\cite{ref:Caldeira&Myhrvold13} (hereafter CM13) fitted three different conceptual models, namely the one-box, two-box and diffusion models to the annual global mean air temperature time series of the CMIP5 \texttt{abrupt4xCO2} experiments \cite{ref:CMIP5}. These fits were then tested against the \texttt{1pcCO2} CMIP5 experiments. CM13 showed that while the one-box model was a poor fit to either experiment on longer timescales both the two-box and diffusion models did good jobs. Here we use these conceptual models to analyse the annual global mean air temperature variability in the CMIP5 \texttt{historical} experiments with a view to obtaining ECS as a function of $\Psi$ as found in CHW18 for the one-box model.

Each of the conceptual models have differing numbers of free parameters, the one-box and diffusion model have three and the two-box has five. None of these are assumed to be fixed. These parameters are essentially fitted to each of the CMIP5 models and the observations in the historical period via $\Psi$ (introduced in equation \ref{eq:EC_ECS}). The models are introduced in order of complexity and completeness, the one-box being the simplest analytically while the diffusion model is harder to solve but reproduces more of the observed temperature response.

The historical time series can be approximated as the sum of the responses to the forcing resulting from changes in the atmospheric composition. These include greenhouse gases, tropospheric and stratospheric aerosols from large volcanic eruptions and solar variability. There is also a response to fast, internal variability which is parameterized here as the response to random, white noise forcing. We seek to isolate the response to the latter and relate it to equilibrium climate sensitivity (ECS). A relation between the system response to random fluctuations and its sensitivity, is essentially a fluctuation-dissipation theorem (FDT) \cite{ref:Kubo66,ref:Leith75}.

ECS is defined as the equilibrium temperature change due to the constant forcing $Q_{2\times CO_2}$ from the doubling of CO$_2$,
\begin{equation}\label{eq:ECSdef}
ECS=\frac{Q_{2\times CO_2}}{\lambda},
\end{equation}
and $\lambda$ is the climate feedback factor.

Linearity of the conceptual models allows each temperature response $T_i(t)$  to each forcing $Q_i(t)$ to be added to give the total response i.e. if the total forcing is $Q(t)=\sum_i Q_i(t)$ then the total temperature response is just the sum of the temperature responses to each of the individual forcings $T(t)=\sum_i T_i(t)$ (principle of superposition). Linearity means that by suitable detrending the response from the trend in emissions can be removed from the total temperature response to leave the residual response, $\Delta T(t)$, to the random forcing. For this study, we assume this detrending can be carried out to a good approximation and work with just the residual temperature. For notational ease we also refer to $\Delta T$ as $T$ i.e. $\Delta T := T$.

Although the theory we derive here assumes external, random forcing, we have shown that the $\Psi$-ECS linear proportionality will theoretically become more tightly defined in the presence of common (non-random) forcing across a model ensemble\cite{ref:Cox18a}. The gradient of the relationship does however change, being roughly inversely proportional to the amplitude of the forcing (see section \ref{sec:CMIP5comp}).

The superposition principle implies the response to any forcing can be written as the convolution of the linear response function $g(t)$ (the response to delta function forcing) with the forcing i.e.
\begin{equation}\label{eq:LRF}
T(t)=\int_0^t g(t-s) Q(s) ds.
\end{equation}
Each model can therefore be characterized by $g(t)$. We will be interested in their response in the stationary limit i.e. when $t \gg \tau$ where $\tau$ is the longest timescale in the model. The residual response is found
when $Q(t)$ is a Gaussian random variable with zero mean and a standard deviation of $\sigma_Q$, $Q(t)=\sigma_Q dW_t$, turning equation \ref{eq:LRF} into a stochastic integral
\begin{equation}\label{eq:LRFstochastic}
T(t)=\sigma_Q \int_0^t g(t-s) dW_s
\end{equation}
where $W_s$ is a Wiener process.

In the following we choose to use the two observables variance $\sigma_T^2$ and autocorrelation $\alpha_T(t)$ as fitting parameters for the conceptual models as they can be easily estimated for a given time series, be it a CMIP5 model or observations. These can be computed for the residual temperature by using equation \ref{eq:LRFstochastic} and the relevant model $g(t)$ via the autocovariance $R(t)$
\begin{align}\label{eq:autocovariance_def}
R(t)&=\lim_{P\rightarrow \infty} \frac{1}{P}\int_0^P T(s-t) T(s) ds.\\
\sigma_T^2&=R(0),\\
\alpha_T(t)&=\frac{R(t)}{R(0)}.
\end{align}

Another useful quantity we use to compare the simple models to the CMIP5 models and observations is the power spectrum of $T$, $|T(\omega)|^2$, which can be found from the Fourier transform of the autocovariance
\begin{equation}\label{eq:ps_to_acf}
|T(\omega)|^2=\int_{-\infty}^\infty R(t) e^{-i\omega t}dt.
\end{equation}

\subsection{Hasselmann one-box model}

The simplest, one-box (or one-timescale) model for the evolution of $T(t)$ is
\begin{equation}\label{eq:oneboxmodel}
C\frac{d T}{dt} = Q(t) - \lambda T
\end{equation}
In this model the climate system can be thought of as a single well-mixed box with effective heat capacity $C$ forced by $Q(t)$ and adjusting to this forcing with climate sensitivity $\lambda$ proportional to the temperature anomaly. The single well-mixed box can be roughly thought of as representing the atmosphere, surface mixed ocean layer and the land.

The linear response function for this model is
\begin{equation}\label{eq:HasselmannLRF}
g(t)=\Theta(t) \frac{e^{-\frac{t}{\tau_H}}}{\lambda \tau_H}
\end{equation}
where the timescale in the model $\tau_H=\tfrac{C}{\lambda}$ and $\Theta(t)$ is the Heaviside step function. When the forcing $Q(t)$ is Gaussian white noise, equation \ref{eq:oneboxmodel} is known as the Hasselmann model \cite{ref:Hasselmann76}. Variance and autocorrelation for the one-box model can be computed from equations \ref{eq:LRFstochastic} and \ref{eq:autocovariance_def} to be
\begin{align}
\sigma_T^2&=\frac{\sigma_Q^2}{2 \lambda^2 \tau_H},\\
\alpha_T(t)&=e^{-\frac{|t|}{\tau_H}}.
\end{align}
These equations can be combined with equation \ref{eq:ECSdef} to give\cite{ref:Cox18}
\begin{equation}\label{eq:EC_ECS}
ECS=\sqrt{2}\frac{Q_{2\times CO_2}}{\sigma_Q}\Psi,
\end{equation}
where $\Psi$ is defined as
\begin{equation}\label{eq:Psi_def}
\Psi=\frac{\sigma_T}{\sqrt{-\log \alpha_{1T}}},
\end{equation}
and $\alpha_{1T}=\alpha_T(1 \text{ year})$. It was equation \ref{eq:EC_ECS}, namely the linear proportionality between the observable $\Psi$, estimated from timeseries of $T$ and the future projection ECS, that was used to guide the search for an emergent constraint in CHW18. The magnitude of proportionality between $\Psi$ and ECS, the ratio of the effective forcing due to doubling CO$_2$ and the mean amplitude of the effective forcing in the experiment $\sigma_Q$, $\sqrt{2}\frac{Q_{2\times CO_2}}{\sigma_Q}$ cannot be observed but is fortunately weakly correlated to ECS ($r=-0.02$) across the CMIP5 model ensemble\cite{ref:Cox18}. By linearly regressing $\Psi$ against ECS, the magnitude of proportionality is therefore determined by the model ensemble itself.

The power spectrum of the one-box model is
\begin{equation}
|T(\omega)|^2=\frac{\sigma_Q^2}{\lambda^2(1+\omega^2 \tau_H^2)}.
\end{equation}
This model predicts a red power spectrum temperature response, that is, the power scales inversely to the square of the forcing frequency $\omega$.

\subsection{Two-box model}\label{sec:two_box}

The two-box model\cite{ref:Gregory00,ref:Held10,ref:Geoffroy13} consists of two well-mixed layers, one representing the upper mixed layer of the ocean plus the lower atmosphere, with effective heat capacity $C$ and temperature $T$, and a second well-mixed box representing the deep ocean with heat capacity $C_0$ and temperature $T_0$. Heat transport between the two boxes is proportional to the temperature difference between the two boxes with constant of proportionality $\gamma$. The equations describing the evolution of temperature are therefore
\begin{align}\label{eq:twoboxmodel}
C\frac{d T}{dt}& = Q(t) - \lambda T - \gamma(T - T_0),\\
C_0\frac{d T_0}{dt} & = \gamma(T - T_0).
\end{align}
This model has a two timescales, a fast $\tau_f$ and slow one $\tau_s$. The linear response function is the sum of the fast and slow modes with amplitudes $\frac{a_f}{\tau_f}$ and $\frac{a_s}{\tau_s}$,
\begin{equation}
g(t)=\frac{\Theta(t)}{\lambda}\left(\frac{a_f}{\tau_f}e^{-\frac{t}{\tau_f}}+\frac{a_s}{\tau_s}e^{-\frac{t}{\tau_s}}\right).
\end{equation}
This model has been extensively used in previous climate applications and here we use the notation and expressions for the amplitudes and timescales in terms of the quantities in equation \ref{eq:twoboxmodel} as given in Geoffroy et al\cite{ref:Geoffroy13}. They also fitted this model to \texttt{abrupt4xCO2} CMIP5 experiments for which they found two widely separated timescales, typical values being $\tau_f\sim 4$ yrs and $\tau_s \sim 250$ yrs while the dimensionless mode parameters, $a_f$ and $a_s$, were roughly of equal size ($a_f \sim 3/5$ and $a_s \sim 2/5$).

The autocovariance function for Gaussian white noise forcing can be found by using equations \ref{eq:LRFstochastic} and \ref{eq:autocovariance_def} and in contrast to the one-box model features two modes:
\begin{equation}
R(t)=\frac{\sigma_Q^2}{2\lambda^2}\left\{\frac{a_f^2}{\tau_f} e^{-\frac{|t|}{\tau_f}} + \frac{a_s^2}{\tau_s} e^{-\frac{|t|}{\tau_s}} + \frac{2 a_f a_s}{\tau_f+\tau_s} \left(e^{-\frac{|t|}{\tau_f}}+e^{-\frac{|t|}{\tau_s}}\right)\right\}
\end{equation}
giving
\begin{align}
\sigma_T^2&=\frac{\sigma_Q^2}{2\lambda^2}\left(\frac{a_f^2}{\tau_f}+\frac{a_s^2}{\tau_s}+\frac{4 a_f a_s}{\tau_f+\tau_s} \right),
\end{align}
\begin{widetext}
\begin{align}
\alpha_T(t)&=\frac{(a_f^2\tau_s  e^{-\frac{|t|}{\tau_f}} + a_s^2\tau_f e^{-\frac{|t|}{\tau_s}}) (\tau_f+\tau_s)  + 2 a_f a_s \tau_f \tau_s (e^{-\frac{|t|}{\tau_f}}+e^{-\frac{|t|}{\tau_s}})} {(a_f^2 \tau_s + a_s^2 \tau_f) (\tau_f+\tau_s) + 4 a_f a_s \tau_f \tau_s}.
\end{align}
\end{widetext}
This general result simplifies for typical fitted parameters to the CMIP5 models\cite{ref:Geoffroy13} as the variance and the autocorrelation are dominated by the fast mode.  These can be approximated in the limit ($\tau_s \gg \tau_f$, $a_s \sim a_f$) by:
\begin{align}
\sigma_T^2&\approx\frac{\sigma_Q^2 a_f^2}{2\lambda^2\tau_f},\\
\alpha_T(t)&\approx e^{-\frac{|t|}{\tau_f}}.
\end{align}
The approximate expressions are therefore very similar to the one-box model for the CMIP5 models. Combining these expressions with the equation for ECS gives
\begin{equation}
ECS=\sqrt{2}\frac{Q_{2\times CO_2}}{\sigma_Q a_f}\Psi
\end{equation}
so that the linear relationship between $\Psi$ and ECS found in the one-box model also approximately holds for the two-box model. The constant of proportionality is however different, it has an extra factor in the denominator $a_f$, which is roughly constant and is approximately $a_f\sim \lambda/(\lambda+\gamma)$ over the CMIP5 model range of parameters. Relative standard deviation in $a_f$ is 13\%. The reason for the approximate equivalence between the ECS relations in one- and two-box models is due to the wide separation in timescales between the two modes fitted to the CMIP5 models. As in the one-box case, the `constant' of proportionality between ECS and $\Psi$, $\sqrt{2}\frac{Q_{2\times CO_2}}{\sigma_Q a_f}$ is weakly correlated to ECS ($r=0.03$) across the CMIP5 models and one can linearly regress $\Psi$ against ECS for a theoretical emergent relationship.

In contrast to the one-box, the two-box power spectrum is
\begin{equation}
|T(\omega)|^2=\frac{\sigma_Q^2}{\lambda^2}\left\{\frac{1+\omega^2(a_f \tau_s+a_s\tau_f)^2}{(1+\omega^2\tau_f^2)(1+\omega^2 \tau_s^2)}\right\}.
\end{equation}
which, depending on the size of terms can give red and $\omega^{-4}$ scaling although when fitted to the CMIP5 models, the spectrum is approximately red.

\subsection{Diffusion equation}

The diffusion equation (or heat equation) model \cite{ref:MacMynowski11,ref:Caldeira&Myhrvold13} consists of a continuous vertical layer, $z\geq 0$, where radiative forcing at the surface ($z=0$) causes heating which is transported down through the water column by diffusion (parameterized by diffusivity $\mathcal{D}$), representing heat uptake by the deep ocean. A mixed-layer surface box has also been added in previous studies to add realism \cite{ref:Oeschger75,ref:Hansen85,ref:Fraedrich04} although here we use just the diffusion equation for simplicity. The model is described by a partial differential equation:
\begin{equation}\label{eq:diffusion_eq}
\frac{\partial T}{\partial t}=\mathcal{D}\frac{\partial^2 T}{\partial z^2}
\end{equation}
with flux boundary conditions
\begin{align}\label{eq:diff_boundary}
-\rho c_p \mathcal{D}&\frac{\partial T}{\partial z}\biggm|_{z=0}=Q(t)-\lambda T(0,t),\\
&\frac{\partial T}{\partial z}\biggm|_{z=z_{max}}=0.
\end{align}
where $\rho$ and $c_p$ are the density and specific heat capacity of water respectively. The maximal depth of the ocean, $z_{max}$, is taken to be infinite. The temperature is now a function of both depth and time, $T(z,t)$ although our interest is only in the surface temperature $T(0,t)$.

In contrast to the one- and two-box models, the ocean is modelled as a vertical continuum rather than a finite number of well-mixed boxes. As heat is diffused down the water column with time, the effective heat capacity increases as the heat sees more ocean. This model can also be thought of as an $M$-box model where $M$ is very large and each well-mixed box is very thin resulting in a continuum of ($M$) timescales. The diffusion model reduces to a one-box model when times of interest are larger than the time taken for heat to be well diffused throughout the water column i.e. when $t > z_{max}^2/\mathcal{D}$. For ocean depths of $z_{max}=4000$ m and typical diffusivities of $\mathcal{D}= 5\times 10^{-5}$ m$^2$ s$^{-1}$ this happens when $t>$ 10,000 years and so this limiting case is not met for the application here.

The linear response function for surface temperature $T(0,t)$ can be found using Laplace transforms on equations \ref{eq:diffusion_eq} and \ref{eq:diff_boundary} to be
\begin{equation}
g(0,t)=\frac{\Theta(t)}{\lambda}\left\{\frac{1}{\sqrt{\pi \tau_D t}}-\frac{e^{\frac{t}{\tau_D}}}{\tau_D}\text{erfc}\left(\sqrt{\frac{t}{\tau_D}}\right)\right\}
\end{equation}
A timescale, $\tau_D$, can be identified in this model as $\tau_D=\tfrac{\rho^2 c_p^2 \mathcal{D}}{\lambda^2}$. $\tau_D\sim 25$ years for the mean value of $\mathcal{D}$ found in CM13\cite{ref:Caldeira&Myhrvold13}. One needs to be aware of an unphysical infinity in $g(0,t)$ at $t=0$ because of the time dependence of the effective heat capacity. At $t=0$ this results in zero effective heat capacity and therefore an infinite response. In reality energy is not absorbed in an infinitely thin surface layer and thus care needs to be taken when calculating at $t=0$.

The power spectrum at the surface can be also be found using either Laplace or Fourier transforms. This is given by
\begin{equation}\label{eq:diff_ps}
|T(0,\omega)|^2=\frac{\sigma_Q^2}{\lambda^2}\left\{\frac{1}{1+\sqrt{2|\omega|\tau_D}+|\omega|\tau_D}\right\}.
\end{equation}
The diffusion model therefore predicts a pink spectrum i.e. power scales inversely proportional to $\omega^{-1}$ in contrast to the red spectra predicted by the one- and two-box models.

To obtain the autocovariance function at $z=0$ we start with the power spectrum and Fourier transform it using equation \ref{eq:ps_to_acf}:
\begin{align}\label{eq:autocovariance_diff}
R(t)=\frac{\sigma_Q^2}{2\lambda^2\tau_D}\left\{e^{-\frac{t}{\tau_D }} \left[\text{erfi}\left(\sqrt{\frac{t}{\tau_D}}\right)+\frac{1}{\pi}\text{E}_1\left(-\frac{t}{\tau_D }\right)\right] - e^{\frac{t}{\tau_D}}\left[\text{erfc}\left(\sqrt{\frac{t}{\tau_D}}\right) - \frac{1}{\pi}\text{E}_1\left(\frac{t}{\tau_D }\right)\right]\right\}
\end{align}
this rather long exact expression can be well approximated more compactly as
\begin{equation}\label{eq:covariance_E1_diff}
R(t)\approx \frac{2\sigma_Q^2}{\pi \lambda^2\tau_D} \text{E}_1\left(\sqrt{\frac{\pi t}{\tau_D}}\right)
\end{equation}
where the exponential integral $\text{E}_1(x)$ is defined as $\text{E}_1(x)=\int_x^\infty \frac{e^{-t}}{t}dt$. Unfortunately $R(t=0)=\sigma_T^2$ is also infinite because of the unphysical zero effective heat capacity. $\sigma_T^2$ can however be approximated by taking a very small but finite time, $t_0$. Starting with equation \ref{eq:covariance_E1_diff} and Taylor expanding the exponential integral to zeroth order around $t=0$ results in
\begin{align}
\sigma_T^2&\approx \frac{2\sigma_Q^2}{\pi \lambda^2\tau_D} \left\{-\gamma_{EM} -\log\left(\sqrt{\frac{\pi t_0}{\tau_D}}\right)\right\}\\
&\approx \frac{c_0 \sigma_Q^2}{\pi \lambda^2 } \tau_D^{\frac{1-c_0}{c_0}}
\end{align}
where $c_0=-2\gamma_{EM} - \log(\pi t_0)$, $\gamma_{EM}\approx 0.577$ is the Euler-Mascheroni constant and in the second line, the approximation $\log x= c_0 x^{\frac{1}{c_0}}- c_0$ has been used. This approximation gets better for larger $c_0$ (smaller $t_0$). So for small $t_0$
\begin{equation}\label{eq:variance_diff}
\sigma_T^2 \rightarrow \frac{c_0 \sigma_Q^2}{\pi \lambda^2 \tau_D}
\end{equation}
and the autocorrelation function is
\begin{equation}\label{eq:autocorrelation_diff}
\alpha_T(t)\approx \frac{2}{c_0} \text{E}_1\left(\sqrt{\frac{\pi t}{\tau_D}}\right)
\end{equation}
which is purely a function of $\tau_D$.

\begin{figure}
\includegraphics[width=\columnwidth]{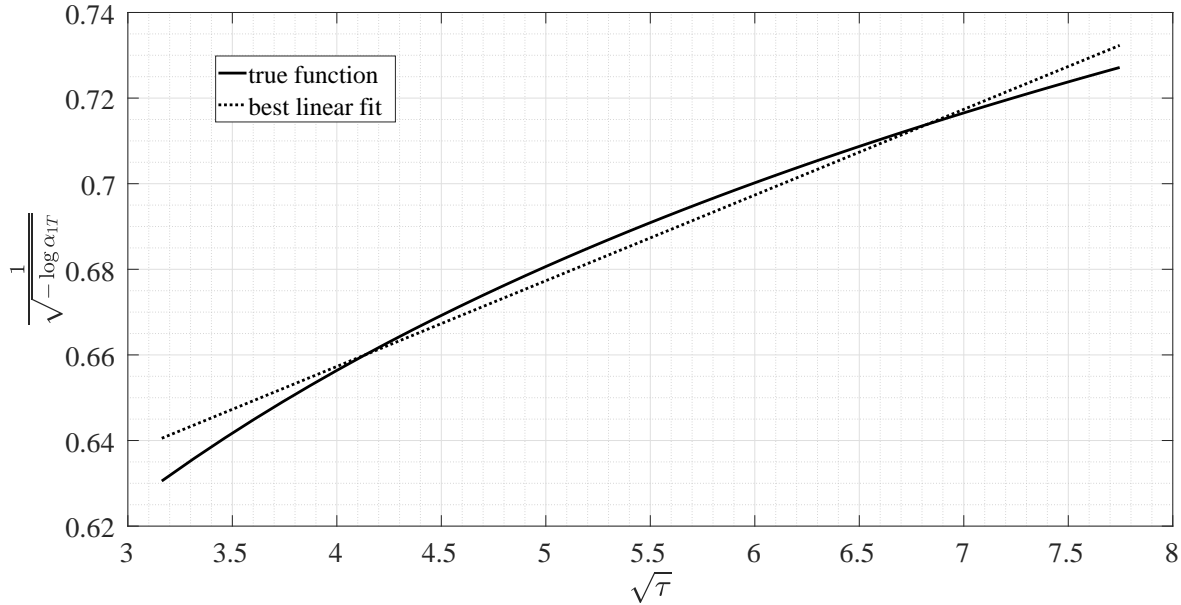}
\caption[]{$\sqrt{\tau_D}$ vs $\frac{1}{\sqrt{-\log \alpha_{1T}}}$ for the diffusion model. If these two functions are linearly proportional then $\Psi$ is also linearly proportional to ECS for the diffusion model. Although it is slightly nonlinear, for this range of values (solid line) linearity seems to be a good approximation (dotted line). $\alpha_{1T}$ is calculated from the exact formula, equation \ref{eq:autocovariance_diff} with $t_0=10^{-6}$ yrs ($\sim 1$ min) and $\tau_D \in [10, 60]$ years. This spans the range of values of $\tau_D$ found in fits to CMIP5 models\cite{ref:Caldeira&Myhrvold13}.}  \label{fig:diff_linear}
\end{figure}

Rearranging equation \ref{eq:variance_diff} and combining with the equation for ECS (eq \ref{eq:ECSdef}) gives
\begin{equation}
ECS=\sqrt{\frac{\pi}{c_0}}\frac{Q_{2\times CO_2}}{\sigma_Q}\Psi_D
\end{equation}
where
\begin{equation}\label{eq:Psi_D}
\Psi_D=\sigma_T\sqrt{\tau_D}.
\end{equation}
Or in terms of observables
\begin{equation}
\Psi_D=\frac{\sigma_T \sqrt{\pi}}{\text{E}_1^{-1}\left(\frac{c_0 \alpha_{1T}}{2}\right)}
\end{equation}
where E$_1^{-1}(x)$ is defined as the inverse of the exponential integral. The linear $\Psi$-ECS proportionality is not true for the diffusion model. However, comparing eq. \ref{eq:Psi_D} with similar for one- and two-box models (eq. \ref{eq:Psi_def}), if
\begin{equation}
\sqrt{\tau_D} \propto \frac{1}{\sqrt{-\log\alpha_{1T}}}
\end{equation}
is approximately true then the linear ECS-$\Psi$ proportionality is also approximately true for the diffusion model. By plotting one against the other in figure \ref{fig:diff_linear} this is the case for the range of values of $\tau_D$ applicable to CMIP5 models ($\tau_D \in [10, 60]$ years in CM13). $\alpha_{1T}$ is calculated from the exact formula, equation \ref{eq:autocovariance_diff}.

\section{Comparison with CMIP5 models and observations}\label{sec:CMIP5comp}

\begin{figure}
\includegraphics[width=\columnwidth]{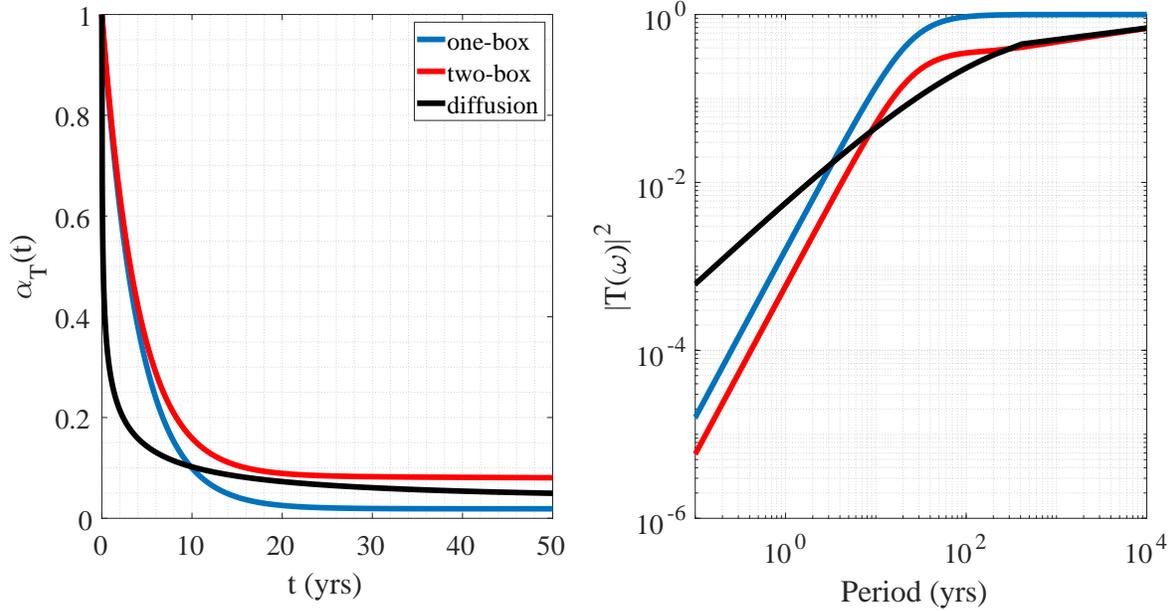}
\caption[]{Autocorrelation, $\alpha_T(t)$, (left) and power spectrum, $|T(\omega)|^2$, (right) for the three conceptual models. $\lambda=\sigma_Q=1$ are the same in all curves, while $\tau=\tau_f=4$ yrs for the one- and two-box models. For the two-box model $\tau_s=250$ yrs and $a_f=3/5$, $a_s=2/5$ (these are the mean values found by Geoffroy et al\cite{ref:Geoffroy13} in fits to the CMIP5 models). For the diffusion model $\tau_D=25$ years. Power spectra and autocorrelation functions are roughly the same for one- and two-box models at timescales less than a decade. For short periods, the diffusion model has a $|T(\omega)|^2 \propto \omega^{-1}$  (power proportional to period) pink spectrum whereas the one- and two-box models show a $|T(\omega)|^2 \propto \omega^{-2}$ (power proportional to the square of the period) red spectra.}\label{fig:simple_model_spectra_acf}
\end{figure}

Theoretical autocorrelation functions and power spectra predicted by the conceptual models are shown in figure \ref{fig:simple_model_spectra_acf} for typical values found in fits to the CMIP5 models\cite{ref:Geoffroy13,ref:Caldeira&Myhrvold13}. One- and two-box autocorrelation functions and power spectra are very similar for timescales less than 100 years. Power spectra in these models have the same $|T(\omega)|^2 \propto \omega^{-2}$ red power spectra. In contrast to the box models, the diffusion model has a faster drop off in autocorrelation but a slower approach to equilibrium and a power spectrum that predicts a $|T(\omega)|^2 \propto \omega^{-1}$ pink power spectrum.

\begin{figure}
\includegraphics[width=\columnwidth]{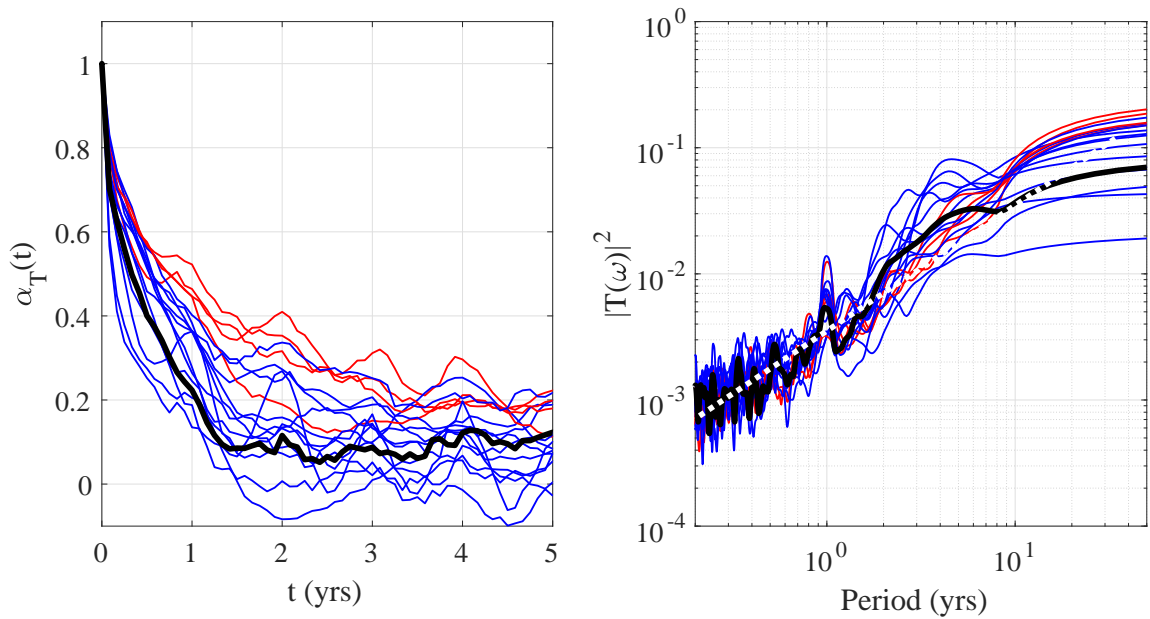}
\caption[]{Autocorrelation, $\alpha_T(t)$, (left) and power spectrum, $|T(\omega)|^2$, (right) for the CMIP5 model \texttt{historical} runs. The CMIP5 models used are the same as in CHW18\cite{ref:Cox18}. Red lines are higher sensitivity models ($\lambda<1$ W m$^{-2}$ K$^{-1}$) while blue lines are lower sensitivity models ($\lambda\geq1$ W m$^{-2}$ K$^{-1}$). The black line is the (historical) HadCRUT4 observational dataset. The white dotted line in the right hand power spectrum plot is a guide to show the $|T(\omega)|^2 \propto \omega^{-1}$ pink spectrum predicted by the diffusion model. The power spectra have been smoothed with a 25 point moving average window.}\label{fig:CMIP5spectra_acf_historical}
\end{figure}

For comparison with the conceptual plots, the CMIP5 \texttt{historical} runs (coloured lines) and the HadCRUT4 historical observational dataset \cite{ref:Morice12} (thick black line) are shown in figure \ref{fig:CMIP5spectra_acf_historical}. The power spectra of the HadCRUT4 observations and CMIP5 models show approximately a $|T(\omega)|^2 \propto \omega^{-1}$ pink spectrum most closely resembled by the diffusion model. The dotted white line is shown as a guide to this proportionality. High sensitivity CMIP5 models also generally have higher autocorrelation. The HadCRUT4 autocorrelation is more representative of the low sensitivity models, consistent with the findings of CHW18.

We have used detrended CMIP5 \texttt{historical} simulations as a comparison to observations can also be made and an emergent constraint obtained. However, conceptual model theoretical relations have been derived assuming white noise external forcing as a parameterization of internal variability. The CMIP5 \texttt{piControl} experiments are the closest analogue to this simplification and one may wonder whether the same relations hold in these experiments as it is known the forced response may not always be the same as the response to internal variability \cite{ref:Lucarini&Sarno11,ref:Gritsun&Lucarini17}. Power spectra and autocorrelation functions for the \texttt{piControl} experiments are broadly the same as figure \ref{fig:CMIP5spectra_acf_historical} (not shown). The linear $\Psi$-ECS emergent relationships are also similarly strong in both \texttt{piControl} and \texttt{historical} simulations having correlations of $r=0.68$ and $r=0.77$ respectively. The higher correlation in the \texttt{historical} experiment resulting in a reduced uncertainty emergent constraint is theoretically expected when there is common forcing across the model ensemble (see Cox et al \cite{ref:Cox18a}). In this case the common forcing in the \texttt{historical} experiment is provided by the increasing concentrations of greenhouse gases, aerosols and volcanic eruptions.

\begin{figure}
%\begin{subfigure}[b]{0.5\textwidth}
\includegraphics[width=\columnwidth]{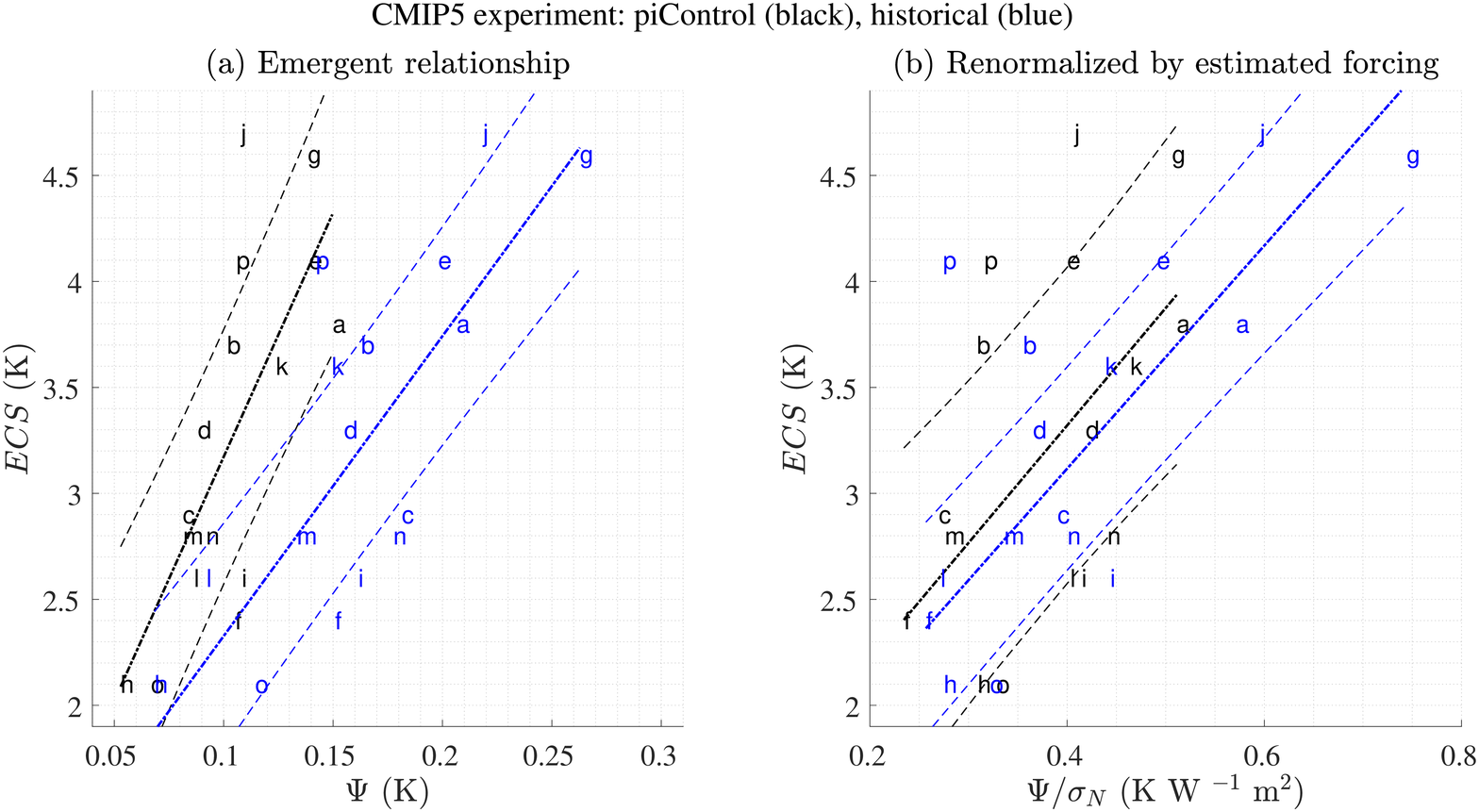}
\caption[]{(a) $ECS$ vs $\Psi$ emergent relationships in the \texttt{historical} (blue) and \texttt{piControl} (black) CMIP5 experiments. Each letter plotted is a CMIP5 model and corresponds to the same used in CHW18 (model-letter correspondence is given in table \ref{table:CMIP5}). The emergent relationship is calculated between 1880-2015 for the \texttt{historical} experiment and for the first 135 years of \texttt{piControl} using the same methodology of CHW18. The gradient of the emergent relationship (dashed line) is theoretically predicted to be smaller with increased forcing in the experiment. This is why the historical run with its volcanic, greenhouse gas, aerosol and internal forcing has a shallower gradient than the control run. (b) is the same plot with $\Psi$ renormalized by the estimated forcing from $\sigma_N$, the standard deviation of the top-of-atmosphere radiative forcing. Emergent relationships then have a very similar gradient illustrating the inverse proportionality of the gradient to the forcing in (a).}\label{fig:emergent_relationship}
%\end{subfigure}
\end{figure}

There are differences in the emergent relationships however. In figure \ref{fig:emergent_relationship} (a) the emergent relationships for the \texttt{piControl} and \texttt{historical} have different gradients. This is due to increased effective forcing in the historical simulations from residual volcanic, aerosol and greenhouse gas forcing remaining after the detrending procedure. From equation \ref{eq:EC_ECS} an inverse relationship with the magnitude of the effective forcing $\sigma_Q$ is expected. When $\Psi$ is divided by the estimated forcing, $\sigma_N \sim \sigma_Q$, in figure \ref{fig:emergent_relationship} (b) gradients are very similar. Forcing has been inferred in the CMIP5 models from the net top-of-atmosphere radiative flux $N$ where $N\approx Q - \lambda T$ with standard deviation $\sigma_N$.

\begin{table}
\caption{CMIP5 models used in figure \ref{fig:emergent_relationship} (letters in the first column identify the models) and in CHW18. Values are taken from IPCC AR5 table 9.5 \cite{ref:IPCC}. %\textbf{The values on the BNU model were done via a different method to the others and look a bit suspect. I get $\lambda=0.93$, $Q_{2\times}=3.72$ via Gregory method. Note that this model is also a bit of an outlier in the emergent relationship plots.
}\label{table:CMIP5}
\begin{tabular}{| c | c | c | c | c |}
  \hline
  % after \\: \hline or \cline{col1-col2} \cline{col3-col4} ...
    & Model         & ECS (K) & $\lambda$ (W m$^{-2}$ K$^{-1}$) & $Q_{2\times CO_2}$ (W m$^{-2}$) \\
    \hline
  a & ACCESS1-0     & 3.8     & 0.8                             & 3.0   \\
  b & CanESM2       & 3.7     & 1.0                             & 3.8   \\
  c & CCSM4         & 2.9     & 1.2                             & 3.6   \\
  d & CNRM-CM5      & 3.3     & 1.1                             & 3.7   \\
  e & CSIRO-MK3-6-0 & 4.1     & 0.6                             & 2.6   \\
  f & GFDL-ESM2M    & 2.4     & 1.4                             & 3.4   \\
  g & HadGEM2-ES    & 4.6     & 0.6                             & 2.9   \\
  h & inmcm4        & 2.1     & 1.4                             & 3.0   \\
  i & IPSL-CM5B-LR  & 2.6     & 1.0                             & 2.7   \\
  j & MIROC-ESM     & 4.7     & 0.9                             & 4.3   \\
  k & MPI-ESM-LR    & 3.6     & 1.1                             & 4.1   \\
  l & MRI-CGCM3     & 2.6     & 1.2                             & 3.2   \\
  m & NorESM1-M     & 2.8     & 1.1                             & 3.1   \\
  n & bcc-csm1-1    & 2.8     & 1.1                             & 3.2   \\
  o & GISS-E2-R     & 2.1     & 1.8                             & 3.8   \\
  p & BNU-ESM       & 4.1     & 1.0                             & 3.9   \\
  \hline
\end{tabular}
\end{table}

\section{Discussion and Conclusions}

All three conceptual models have both physical similarities and deficiencies relative to the CMIP5 models and the real Earth system. The one-box model only really has any physical justification when the timescales of interest are dominated by the well-mixed atmosphere and ocean surface layer. This has led some to question the use of the one-box model by CHW18 to motivate their emergent constraint between equilibrium climate sensitivity (ECS) and $\Psi$, a statistic dominated by the fast timescale processes e.g. Rypdal et al\cite{ref:Rypdal18}. However, in this paper we have shown that a near-linear relationship is to be expected between ECS and $\Psi$ for more realistic conceptual models. For the one- and two-box models we were able to find analytical relations to show this. Semi-analytical relations for the diffusion model also show a similar near linear relationship.

Even though a linear proportionality between $\Psi$ and ECS is expected in the conceptual models for regions of parameter space applicable to CMIP5 models, each of the conceptual models predicts different temperature responses. The one-box model cannot reproduce the long timescale behaviour of the two-box or diffusion model and neither the one- or two-box models can mimic the observed and CMIP5 power spectra. Of the three, the diffusion model reproduces the power spectra of the CMIP5 models and the observations most closely although it is more difficult to work with and has some deficiencies as an analogue to the real climate system. Combining a well-mixed atmosphere-surface ocean box with a diffusive continuous deep ocean\cite{ref:Oeschger75,ref:Hansen85,ref:Fraedrich04}, although adding another layer of complexity and making the model less analytically amenable, would add physical realism. We suspect this would produce a similar linear relation to the one- and two-box models as well as mimicking the CMIP5 and HadCRUT4 power spectra due to the timescale separation between surface mixed and deep layers.

In conceptual models, we therefore expect to find emergent relationships between ECS and short-term variability (e.g. as measured by $\Psi$). However, the underlying models considered here remain deliberately very simple compared to the GCMs we are using to define emergent constraints. It is therefore vital that we continue to check that our conceptual models provide useful insights into the spread of projections from GCMs. We see this as a form of hypothesis testing, in which a conceptual model is proposed, an emergent relationship between variability and sensitivity is predicted based-on that conceptual model, and then that predicted emergent relationship is checked against the ensemble of full-form GCMs. This approach requires that the search for emergent constraints becomes more theory-led than it has been to date, but would also guard against spurious relationships that could easily arise from blind data-mining of the many diagnostics available from modern GCMs. Most importantly, in our view, such theory-led hypothesis testing is much more likely to improve understanding of the climate system than purely-statistically-derived emergent constraints.

\begin{acknowledgments}
This work was supported by the EU Horizon 2020 Research Programme CRESCENDO project, grant agreement number 641816 (P.M.C. and M.S.W.); the EPSRC-funded ReCoVER project (M.S.W.); the European Research Council (ERC) ECCLES project, grant agreement number 742472 (P.M.C. and F.J.M.M.N.); We also acknowledge the World Climate Research Programme’s Working Group on Coupled Modelling, which is responsible for CMIP, and we thank the climate modelling groups for producing and making available their model output.
\end{acknowledgments}

\bibliography{../../bibliography/masterbib}

\end{document}